\documentclass[
showpacs,preprintnumbers,
nofootinbib,
nobibnotes,
amsmath,amssymb,
aps,
floatfix,
]{revtex4-1}

\usepackage{graphicx}
\usepackage{dcolumn}
\usepackage{bm}
\usepackage[colorlinks,linkcolor=blue,urlcolor=blue,
   citecolor=blue,]{hyperref}
\usepackage{multirow,wasysym}
\usepackage{rotating}
\usepackage{braket,mathtools}
\usepackage[	
margin=0.9in,
]{geometry}
\usepackage[subrefformat=parens,labelformat=parens,caption=false]{subfig}
\usepackage{bibentry}
\usepackage{lipsum}

\begin{document}


\title{
Proton impact on ground and excited states of atomic hydrogen 
}
\author{Anthony C. K. \surname{Leung}}
 \email{leungant@yorku.ca}
\author{Tom \surname{Kirchner}}%
 \email{tomk@yorku.ca}

\affiliation{
 Department of Physics and Astronomy, York University, Toronto, Ontario,
 M3J 1P3, Canada
}%

\date{\today}

\begin{abstract}
The processes of electron excitation, capture, and ionization
were investigated in proton collisions with atomic hydrogen 
in the initial $n=1$ and $n=2$
states at impact energies from 1 to 300 keV. 
The theoretical analysis is based on the close-coupling
two-center basis generator method in the semiclassical approximation.
Calculated cross sections are compared with previous results
which include data obtained from classical-trajectory Monte Carlo, 
convergent close-coupling, and other two-center atomic
orbital expansion approaches.
There is an overall good agreement in the capture and excitation
cross sections while there are some discrepancies in the ionization results
at certain impact energies.
These discrepancies in the present results can be partially 
understood through the use 
of a $1/n^{3}$ scaling model.
\end{abstract}

\maketitle

%
%
\section{Introduction}
\label{intro}
The classic problem of proton scattering from ground-state
hydrogen has often been used as a benchmark system
for theoretical models
\cite{Olson1977,Koakowska1998,Toshima1999,Winter2009,Avazbaev2016}. 
Cross sections of electronic processes (e.g., capture, 
excitation, and ionization) for this 
prototypical system have important applications in plasma physics.
In recent times, there is much interest from the
International Thermonuclear Experimental Reactor (ITER)
fusion energy research community \cite{Hemsworth2009} in ion collisions
with initially excited hydrogen atoms.
The International Atomic Energy Agency (IAEA) 
Coordinated Research Program on Data for Atomic
Processes of Neutral Beams in Fusion Plasma aims to provide
recommended data to the ITER project for plasma modeling \cite{Chung2017}.

Theoretical efforts using classical and semiclassical
approaches have been made to obtain cross sections
for proton collisions with excited hydrogen atoms.
Previously, Pindzola et al. \cite{Pindzola2005} performed calculations
for capture and excitation cross sections
for $p$--H($2s$) collisions at impact energies from 1 to 100 keV by
using the classical-trajectory Monte Carlo (CTMC), the
two-center atomic-orbital close-coupling with pseudostates (AOCC-PS),
and the time-dependent lattice (TDL) approaches.
Recent work based on the wave packet convergent
close-coupling (WP-CCC) method \cite{Abdurakhmanov2018} 
and a two-center atomic-orbital close-coupling with Gaussian-type orbitals (AOCC-GTO)
calculation \cite{Agueny2019}
reported similar analyses and also 
examined $p$--H($2p_{0}$) and $p$--H($2p_{1}$) collisions. 
Comparisons of the capture cross sections of 
$p$--H($2s$) collisions from these calculations 
all showed good agreement but there are discrepancies
in the excitation results from 10 to 100 keV impact energy.  
These discrepancies could be due to differences in the
target basis size which has been larger 
in the recent works \cite{Abdurakhmanov2018, Agueny2019} 
than in the AOCC-PS analysis \cite{Pindzola2005}.
These differences could indicate that these additional states serve as 
intermediate channels during the collision.
Ionization cross sections from the WP-CCC calculations were
also reported but no comparisons were made since no other
data were available at the time.

The purpose of the present work is to address the 
need for additional independent analyses of proton collisions
with hydrogen atoms for the IAEA
Coordinated Research Program to help establish the range
of validity of cross section data.
The approach for the present theoretical analysis is  
the semiclassical, nonperturbative 
two-center basis generator method (TC-BGM) \cite{Zapukhlyak2005}. 
It is a close-coupling approach similar to the
AOCC method, but the main feature of the TC-BGM is 
its use of a dynamic basis that is adapted to the problem at hand.
This has the practical advantage that fewer
pseudostates need to be employed to reach convergence 
compared to using the standard approach. 
In this work, the focus is on proton collisions
with atomic hydrogen in the initial $n=1$ and $n=2$ states
at impact energies from 1 to 300 keV. 
This is the region where the discrepancies of excitation 
and ionization cross sections
are largest based on previous comparisons \cite{Abdurakhmanov2018}. 
It is the aim of this study to provide some validation
of the existing results.

The article is organized as follows. In Sect. \ref{sec:theory}
the TC-BGM is outlined. The collision cross section results
are presented and discussed in Sect. \ref{sec:results}.
Finally, concluding remarks are provided in Sect. \ref{sec:end}.
Atomic units ($\hbar=e=m_{e}=4\pi\epsilon_{0}=1$) are used
throughout the article unless stated otherwise.

\section{Two-center basis generator method}
\label{sec:theory}

\begin{figure}[hbtp]
\centering
\includegraphics[scale=1]{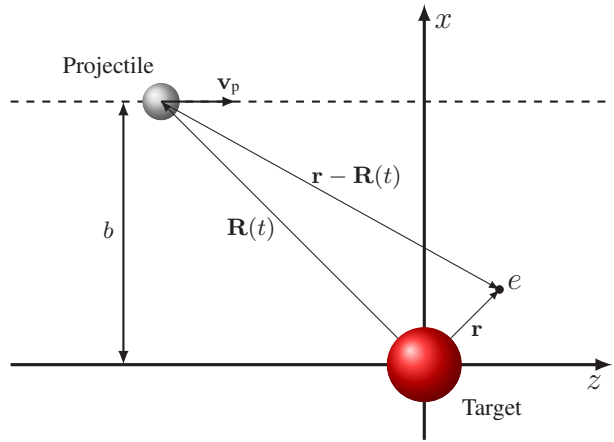}
\caption{Setup of the collision system in the impact parameter model.}
\label{fig:collision}
\end{figure}

The present treatment of the proton-hydrogen collision problem
starts with the impact-parameter model within the semiclassical
approximation. 
Figure \ref{fig:collision} shows the setup of the collision
framework where the $xz$--plane is chosen as the
scattering plane.
In the laboratory frame, 
the hydrogen atom is assumed to be fixed in space 
and the proton travels in a straight-line path at constant speed $v_{\text{P}}$,
described by $\mathbf{R}(t)=(b,0,v_{\text{P}}t)$, where $b$ is the
impact parameter.

In a one-electron collision system, the objective is 
to solve a set of single-particle time-dependent Schr\"{o}dinger equations (TDSEs)
for the initially occupied ground and excited states under consideration,
\begin{equation}\label{eq:TDSEs}
	i{\partial\over\partial t}\psi_{j}(\mathbf{r},t)=\hat{h}(t)\psi_{j}(\mathbf{r},t),~j=1,...,K.
\end{equation}
For the proton-hydrogen collision system,
the target is described by the undisturbed Hamiltonian $\hat{h}_{0}$
which contains the electronic kinetic energy and the Coulomb
potential of the target nucleus.
Let $\hat{V}(t)$ be the time-dependent Coulomb potential
of the proton projectile. The Hamiltonian of the collision system is
\begin{equation}
\begin{split}
	\hat{h}(t) &=\hat{h}_{0} +\hat{V}(t)\\
	&= -{1\over 2}\nabla^{2}-{1\over r}-{1\over |\mathbf{r}-\mathbf{R}(t)|}.
\end{split}
\end{equation}
The present calculation solves the set of 
single-particle TDSEs \eqref{eq:TDSEs} by
the close-coupling approach subject to
all bound states of the hydrogen target in the $n=1$ and $n=2$ shells
as initially occupied states.

The idea of the close-coupling TC-BGM is to expand
the electronic solutions $\psi_{j}(\mathbf{r},t)$ in a basis which
dynamically adapts to the time-dependent problem.
In an early foundational work of the single-center BGM \cite{Kroneisen1999}, 
it was argued that the model space constructed by repeated
application of a regularized Coulombic projectile
potential onto atomic target eigenstates provides such a
dynamical representation of $\psi_{j}(\mathbf{r},t)$. 
Calculations in a two-center framework
are naturally performed in the center-of-mass frame.
The basis is then generated from a finite set of
$N_{\text{T}}$ target and $N-N_{\text{T}}$ projectile atomic states taking
into account Galilean invariance by the appropriate choice
of electron translation factors
\begin{equation}\label{eq:tc-states}
\phi^{0}_{\nu}(\mathbf{r})=
\begin{cases}
	\phi_{\nu}(\mathbf{r}_{\text{T}})\exp{(i\mathbf{v}_{\text{T}}\cdot\mathbf{r})},~~\nu\leq N_{\text{T}}\\
	\phi_{\nu}(\mathbf{r}_{\text{P}})\exp{(i\mathbf{v}_{\text{P}}\cdot\mathbf{r})},~~\text{otherwise,}
\end{cases}
\end{equation}
where $\mathbf{v}_{\text{T}}$ and $\mathbf{v}_{\text{P}}$ are the 
velocities of the target and projectile frames
with respect to the center-of-mass frame, respectively.
In the TC-BGM, the generating two-center atomic basis \eqref{eq:tc-states}
is augmented by BGM pseudostates, which are constructed
by repeated application of a regularized potential
onto the target states only,
\begin{equation}\label{eq:bgm-states}
	\chi^{\mu}_{\nu}(\mathbf{r},t)=\left[W_{\text{P}}(t)\right]^{\mu}\phi^{0}_{\nu}(\mathbf{r}),~~~\mu=1,...,M,
\end{equation}
\begin{equation}
	W_{\text{P}}(t)={1-\exp{[-\alpha r_{\text{P}}(t)]}\over r_{\text{P}}(t)},		
\end{equation}
where $\alpha=1$ is used in practice.
The construction of Eq. \eqref{eq:bgm-states} is what gives
the basis of $\psi_{j}$ a dynamical feature.
The set of pseudostates \eqref{eq:bgm-states}, when orthogonalized
to the generating basis \eqref{eq:tc-states}, 
accounts for quasimolecular effects
at low impact energies and for ionization channels. 

In terms of the basis states $\chi_{\nu}^{\mu}$, 
the single-particle solution for the
$j$-th initial condition is represented as
\begin{equation}\label{eq:single-sol}
\begin{split}
	\psi_{j}(\mathbf{r},t)&=
		\sum_{\mu=0}^{M(\nu)}\sum_{\nu=1}^{N}c^{j}_{\mu\nu}(t)\chi^{\mu}_{\nu}(\mathbf{r},t),\\
M(\nu)&=
\begin{cases}
M & ~~~~\text{if}~\nu\leq N_{\text{T}}\\
0 & ~~~~\text{otherwise}.
\end{cases}
\end{split}
\end{equation}
Substituting Eq. \eqref{eq:single-sol} into Eq. \eqref{eq:TDSEs}
and projecting onto the BGM basis states results
in a set of close-coupling differential equations
which can be expressed in matrix-vector form
\begin{equation}\label{eq:mat-form}
	i\mathsf{S}{d\over dt}\mathbf{c} = \mathsf{M}\mathbf{c},
\end{equation}
where $\mathbf{c}$ is a column vector 
with the expansion coefficients as components, 
$\mathsf{S}$ is the overlap matrix and $\mathsf{M}$ is the interaction matrix.
Equation \eqref{eq:mat-form} can be solved by standard methods.
Probabilities of electronic transitions at given impact parameter
and speed are obtained
from the expansion coefficients in the asymptotic region
\begin{equation}
 p^{j}_{\mu\nu} = \lim_{t\rightarrow\infty}|c^{j}_{\mu\nu}(t)|^{2}.
\end{equation}
Specifically, bound-state probabilities for finding the
electron on the target $p^{\text{tar}}$ or on the
projectile $p^{\text{cap}}$ are calculated from summing up the
transition probabilities within the generating basis \eqref{eq:tc-states}, and
probabilities for total ionization $p^{\text{ion}}$ are obtained
from the unitarity criterion
\begin{equation}\label{eq:ion}
p^{\text{ion}}= 1- p^{\text{tar}}-p^{\text{cap}}.
\end{equation}
Cross sections for the electronic transitions 
are obtained by integrating the probabilities over the impact
parameter
\begin{equation}\label{eq:cross}
\sigma = 2\pi\int_{0}^{b_{\text{max}}}bp(b) db,
\end{equation}
where $b_{\text{max}}$ is the upper bound at which
the integral is cut in practice.
It should be noted that the conservation of unitarity \eqref{eq:ion}
was monitored in the present analysis and it was found that 
deviations produced by the calculations are typically no larger than 1\%.

In the present analysis, 
the basis set $\lbrace \chi_{\nu}^{\mu}\rbrace$ includes
all $nlm$ hydrogen states for $n\in [0, 6]$ on both
the target and projectile.
The basis also includes a set of BGM pseudostates
up to order $\mu=3$.
Table \ref{tab:converge} shows an example of results from
a convergence test for the $p$--H($2s$) system 
at $E_{\text{P}}=15$ keV where the number of BGM pseudostates
was systematically increased until the percent differences
of cross section results are about 1\% or less.
Furthermore, four initial states of 
the hydrogen target $\lbrace 1s, 2s, 2p_{0}, 2p_{1}\rbrace$
are considered in the present calculation.
It is noteworthy that probabilities from collision calculations
for the $2p_{1}$ and $2p_{-1}$ initial state 
are identical due to the symmetry of the collision system.
Propagation of the set of TDSEs
was carried out from $z_{i}=v_{\text{P}}t_{i}=-100$ 
to $z_{f}=v_{\text{P}}t_{f}=100$ a.u.
for each impact parameter from $b_{\text{min}}=0.3$ to $b_{\text{max}}=70$ a.u.
Note that the integration of Eq. \eqref{eq:cross} is still performed
over the interval $[0, b_{\text{max}}]$.
Preliminary calculations showed that excitation from
initially excited states to final target states of 
higher angular momentum decays slowly with increasing impact parameter. 
Therefore, the choice of $b_{\text{max}}=70 $ a.u. 
in the present calculation was
necessary to capture the asymptotic profile of these transition
probabilities with sufficient accuracy.

\begin{table}[htbp]
\caption{Total cross sections ($10^{-16}$ cm\textsuperscript{2}) 
for $p$--H($2s$) collisions at
$E_{\text{P}}=15$ keV with different number of BGM pseudostates.
}
\label{tab:converge}       
\centering
\begin{tabular}{lrr}
\hline\noalign{\smallskip}
Total pseudostates & $\sigma^{\text{cap}}$ & $\sigma^{\text{ion}}$  \\
\noalign{\smallskip}\hline\noalign{\smallskip}
126& 6.96 & 18.62 \\
175 & 6.62 & 20.72 \\
192 & 6.57 & 21.02 \\
203 & 6.58 & 21.24 \\
\noalign{\smallskip}\hline
\end{tabular}

\end{table}

\section{Results and discussion}
\label{sec:results}

To alleviate the discussion of the results, only the main
findings and comparisons are highlighted in this section. 
Detailed results from this analysis such as state-selective
capture are available
from the authors upon request.


\begin{figure*}[hbtp]
\centering
\includegraphics[scale=1]{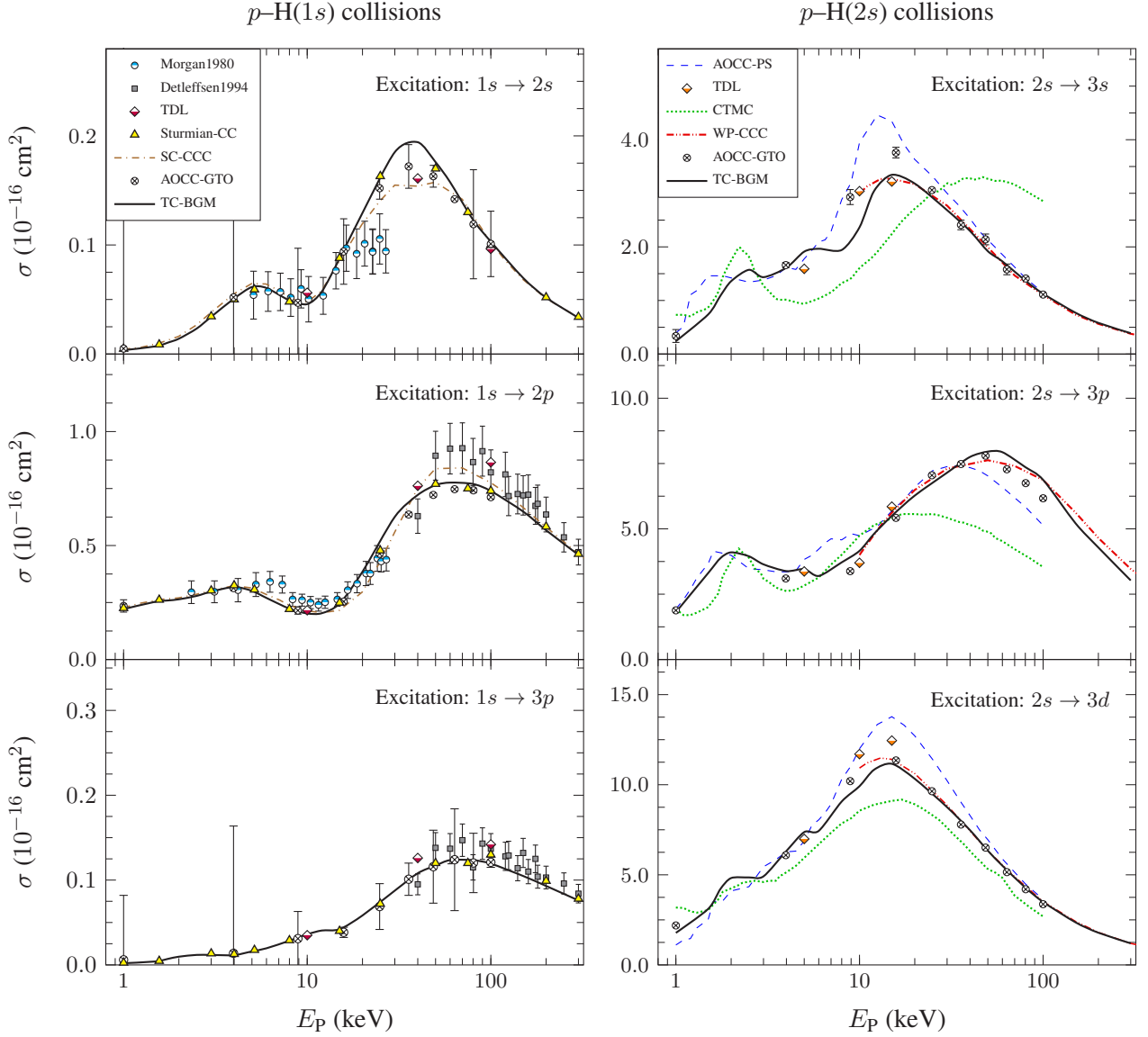}
\caption{State-selective excitation cross sections plotted with respect
to the impact energy for $p$--H($1s$) (left column) 
and $p$--H($2s$) collisions (right column).
Experiment: Morgan1987 \cite{Morgan1980} and Detleffsen1994 \cite{Detleffsen1994}.
Theory: AOCC-PS, CTMC and TDL \cite{Pindzola2005}; 
Sturmian-CC \cite{Winter2009}; SC-CCC \cite{Avazbaev2016}; 
WP-CCC \cite{Abdurakhmanov2018}; AOCC-GTO \cite{Agueny2019};
and present TC-BGM. 
}
\label{fig:total-excite}
\end{figure*}

Results of the $nl$ excitation cross sections
for $p$--H($1s$) and $p$--H($2s$) collisions
are displayed in Fig. \ref{fig:total-excite}.
For H($1s)$ collisions, the present cross sections are compared with
several previous theoretical results based on the
TDL technique \cite{Koakowska1998, Pindzola2005},
close-coupling approaches based on a large Sturmian 
basis (Sturmian-CC) \cite{Winter2009}, 
semiclassical convergent close-coupling (SC-CCC) 
calculations \cite{Avazbaev2016},
and the AOCC-GTO analysis \cite{Agueny2019}.
Note that the uncertainty bars shown from the AOCC-GTO results 
are estimates of the convergence of the cross sections \cite{Agueny2019}.
Previous experimental results of 
Refs. \cite{Morgan1980, Detleffsen1994} are shown alongside.
For H($2s$) collisions, the present cross sections are shown together with
the aforementioned works based on the AOCC-PS \cite{Pindzola2005}, CTMC \cite{Pindzola2005}, WP-CCC \cite{Abdurakhmanov2018} and
AOCC-GTO \cite{Agueny2019} calculations.
Although many earlier works on $p$-H($1s$) collisions 
exist in the literature such as the AOCC calculations of 
Refs. \cite{Kuang1996, Kuang1996a}, the present comparison focuses
on theoretical works within the past decade.

Examining the present excitation results for H($1s)$ collisions,
the cross section profile for all three transitions
exhibits a similar feature where the cross section peaks
at an impact energy between 10 and 100 keV. 
Quantitatively, the $1s\rightarrow 2p$ transition
is largest compared to the other two transitions.
Furthermore, the present TC-BGM results are in excellent agreement
with the Sturmian-CC results \cite{Winter2009} for all three transitions.
Similar agreement can also be seen with other calculations except
around 40 keV where discrepancies are more pronounced.
Although there are also some quantitative discrepancies with the
experimental data \cite{Morgan1980, Detleffsen1994},
for example between 50 and 100 keV for the $1s\rightarrow 2p$ transition, the
present results are mostly within the uncertainty range.

Similar observations can be made for the excitation
cross sections for H($2s$) collisions. 
Specifically, these cross sections also peak 
in the energy region between 10 and 100 keV. 
However, the excitation cross sections for H($2s$) collisions
are larger in magnitude than those of H($1s$) collisions.
From the comparisons with previous results,
it can be seen that the present TC-BGM calculations from 1 to 10 keV
are mostly consistent with the AOCC-GTO \cite{Agueny2019}, AOCC-PS and TDL results
\cite{Pindzola2005} and follow the WP-CCC \cite{Abdurakhmanov2018} trends
at higher energies while the discrepancies with AOCC-PS are more
pronounced in this regime.
Although the cross section curves from the CTMC calculations \cite{Pindzola2005}
have similar structures as the other results, quantitative 
differences in the intermediate region are apparent.
Moreover, although both the present TC-BGM and the AOCC approaches utilize
a basis-set expansion technique in the semiclassical approximation, 
the discrepancies between the TC-BGM and both AOCC data sets \cite{Pindzola2005,Agueny2019} 
may be attributed to the differences in the basis.
Both AOCC-PS \cite{Pindzola2005} and AOCC-GTO \cite{Agueny2019}
calculations included bound states of hydrogen 
with angular momenta up to $l=4$ compared
to $l=5$ in the present calculations.
However, the AOCC-GTO calculation has more pseudostates included
in the basis than the AOCC-PS calculation, which may explain
the discrepancies between those results.
Additional TC-BGM calculations also
showed that a reduced basis on the target, for example including all 
$nlm$ states up to $n=5$ only, had larger discrepancies
with the WP-CCC results at 15 keV.
However, such a change of the basis does not necessarily 
affect the excitation cross sections at all
impact energies.

\begin{figure*}[hbtp]
\centering
\includegraphics[scale=1]{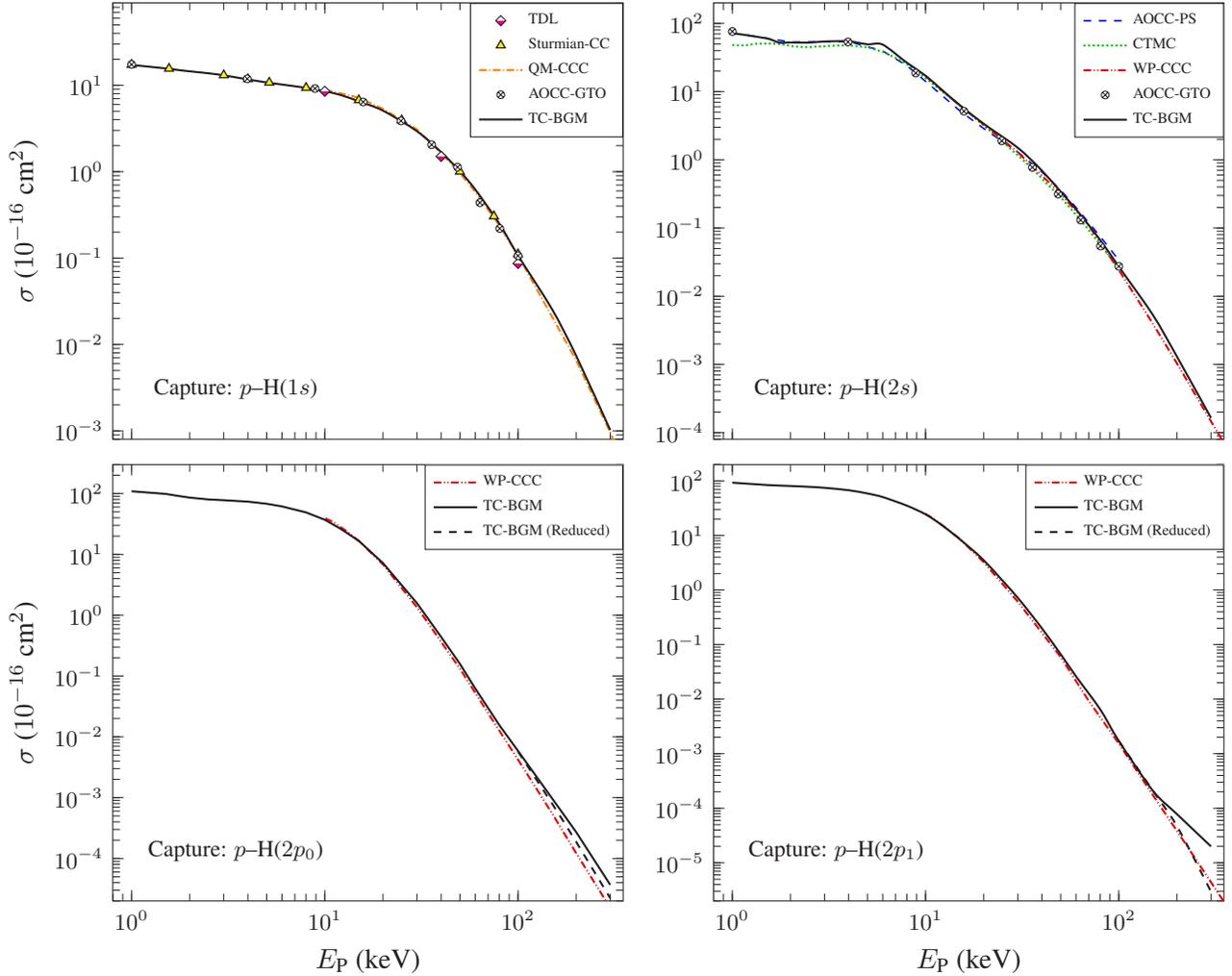}
\caption{Total capture cross sections for the proton--hydrogen
collision problem.
Calculations include: TDL \cite{Koakowska1998};
AOCC-PS and CTMC \cite{Pindzola2005}; 
Sturmian-CC \cite{Winter2009}; QM-CCC \cite{Abdurakhmanov2016};
WP-CCC \cite{Abdurakhmanov2018}; AOCC-GTO \cite{Agueny2019}; and present TC-BGM.
}
\label{fig:total-cap}
\end{figure*}

Figure \ref{fig:total-cap} shows the total
capture cross sections plotted with respect to the impact 
energy for the $p$-H scattering system.
Calculations for the four initial states of the
hydrogen target $\lbrace 1s, 2s, 2p_{0}, 2p_{1}\rbrace$
are shown in separate plots.
Theoretical results from the aforementioned 
studies \cite{Koakowska1998, Pindzola2005,Winter2009,Abdurakhmanov2018}
are shown alongside the present results.
For H($1s$) collisions the included results from Ref. \cite{Abdurakhmanov2016} 
are from calculations based on
the quantum-mechanical convergent close-coupling (QM-CCC) method,
which is viewed as an exact treatment of the quantum-mechanical
three-body Schr\"{o}dinger equation.
It is noteworthy that the AOCC-GTO calculation \cite{Agueny2019}
considered all $2p_{m}$ initial states, 
but the reported cross sections are 
averaged over the $m$ substates, and thus, not appropriate to
compare with the results shown in the present figure.
Comparisons of the present TC-BGM capture cross sections for
H($2p$) collisions, when averaged across the $m$ states,
agree with the AOCC-GTO calculation within 2\% or better.

The present capture results show the expected
fall-off with increasing energy.
At low energies from 1 to 10 keV, the cross sections
for H($n=2$) collisions are larger than those 
of H($n=1$) collisions.
In general, the present results are in very good
agreement with previous calculations \cite{Koakowska1998, Winter2009, Abdurakhmanov2016, Abdurakhmanov2018}.
Although there is good qualitative agreement
between the present results and
the WP-CCC calculations for the
$p$--H($2p_{0}$) and $p$--H($2p_{1}$) systems,
quantitative discrepancies are apparent at 100 keV and higher.
Further investigation showed that the TC-BGM capture into the
$n=6$ shell is about an order of magnitude larger
than the capture into lower shells at these energies, which appears
to indicate a numerical precision issue.
This is evident when capture channels
from higher $n$-states that do not follow the expected distribution 
are excluded from the sum as shown
in Fig. \ref{fig:total-cap} as ``TC-BGM (Reduced)'', 
resulting in a total cross section that is
closer to the WP-CCC result.
While in principle these cross section can be 
improved by fine-tuning numerical parameters
in the calculations this 
is rather cumbersome for this problem
since, for example, slightly reducing the grid points leads
to convergence problems while a slightly denser
grid creates a significant increase in computation time.
In view of these difficulties, no adjustments were made
to improve the numerical accuracy of the calculations at high energies.

\begin{figure}[hbtp]
\centering
\includegraphics[scale=0.98]{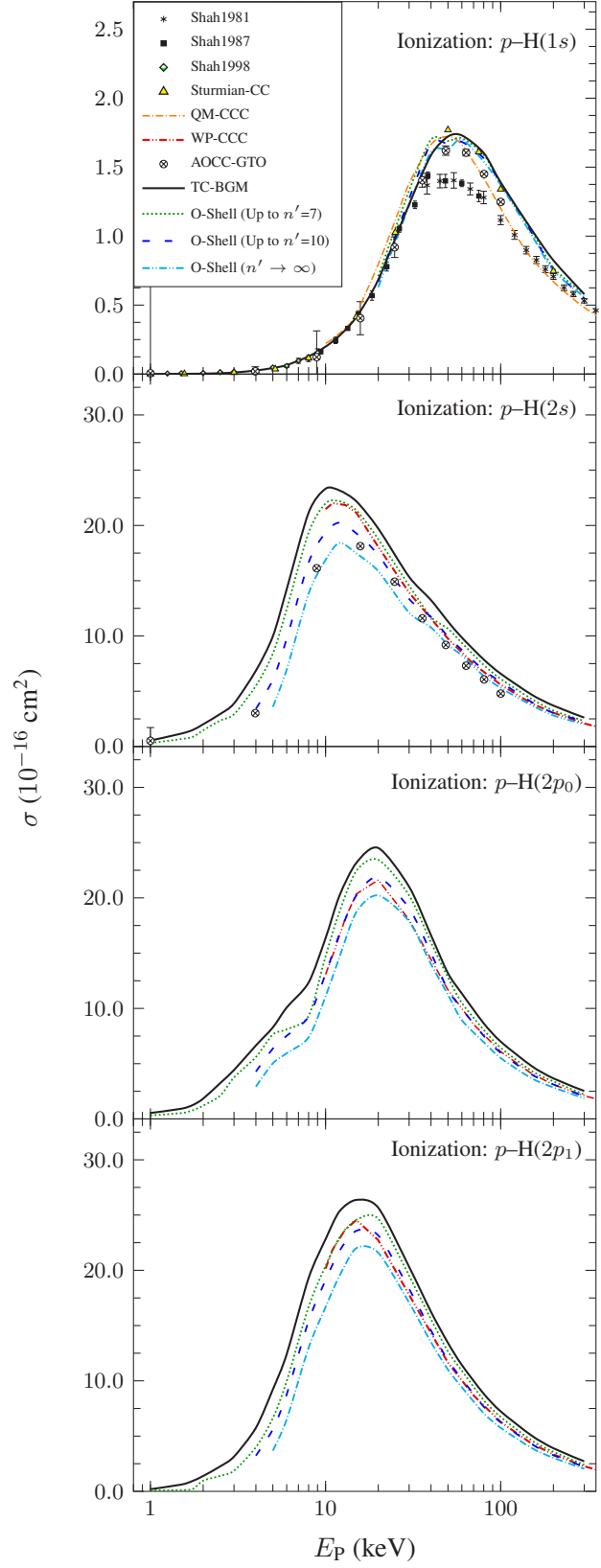}
\caption{Ionization cross sections plotted with respect
to the impact energy for $p$--H collisions.
Experiment: Shah1981 \cite{Shah1981}; Shah1987 \cite{Shah1987}; and Shah1998 \cite{Shah1998}. 
Theory: QM-CCC \cite{Abdurakhmanov2016}; WP-CCC \cite{Abdurakhmanov2018};
AOCC-GTO \cite{Agueny2019};
and present TC-BGM. 
}
\label{fig:total-ion}
\end{figure}

In Fig. \ref{fig:total-ion}, several sets of ionization cross sections
for H($n=1$) and H($n=2$) collisions are compared with the present TC-BGM results.
The sets include experimental measurements
by Shah and co-workers \cite{Shah1981, Shah1987, Shah1998} 
along with theoretical data from 
QM-CCC \cite{Abdurakhmanov2016}, WP-CCC \cite{Abdurakhmanov2018}, 
and AOCC-GTO \cite{Agueny2019} approaches.
Additional results based on a scaling model are also presented
for comparison in order to help understand some of the
discrepancies shown in the present results. 
The use of this scaling model is explained as follows.

The well-known $1/n^{3}$ scaling law 
from perturbation theory \cite{Omidvar1967} was applied to the present results
to extrapolate transitions to
higher $n$-shells. For capture, if $p_{n}$ is the probability
of transfer into the $n$-th shell of the projectile an assumed $1/n^{3}$
scaling predicts that
\begin{equation}\label{eq:scale}
	p_{n+k}=\left(n\over n+k\right)^{3}p_{n},~~k=1,2,...
\end{equation}
is the probability of transfer into the $(n+k)$-shell 
of the projectile.
A detailed analysis showed that using the present results
for capture into $n=3,~4,~\text{or}~5$
to predict the capture into higher shells 
approximately fulfilled this scaling model.
Although Eq. \eqref{eq:scale} applies to electron capture \cite{Omidvar1967}
the present analysis suggests that the excitation 
results also approximately follow this scaling law.
The exception to Eq. \eqref{eq:scale} 
is capture from H($1s$) at low energies since the process
is highly selective to a particular $n$-shell and 
the drop-off in higher shells does not follow the $1/n^{3}$ scaling.
Moreover, the analysis showed that using the $O$-shell
as the reference point where $n=5$ is fixed in Eq. \eqref{eq:scale}
yielded probabilities for higher $n$-shells 
that are modest in magnitude (i.e., neither too small nor too large)
compared to using other reference shells.
This scaling model in the present analysis is referred to as
the $O$-shell model. 
With probabilities of the electronic transitions
to higher $n$-shells computed in this way, new ionization
probabilities are obtained from the unitarity criterion \eqref{eq:ion}.
By denoting $n'=n+k$, several choices of $n'$ were made
to obtain additional sets of ionization cross sections.
As shown in Fig. \ref{fig:total-ion}, states up to $n'=7$ 
were mainly considered in the low-energy region
while states up to $n'=10$ were considered in the
intermediate region. 
The extreme case of $n'\rightarrow\infty$,
which in practice is computed up to $k=200$,
is also included for comparison.
Note that the scaling results for the latter two cases
are not shown in the low-energy region since 
large sums of Eq. \eqref{eq:scale}
turned out to violate the unitarity criterion \eqref{eq:ion}
at those energies.

It is expected that ionization is significant
in the intermediate energy regime between 10 keV and 1 MeV.
Starting with ionization from $p$--H($1s$) collisions, 
the present TC-BGM cross sections reflect this behavior.
In terms of comparisons, the
present TC-BGM results are in good agreement with 
the Sturmian-CC results \cite{Winter2009}.  
Although the present cross sections are consistent with the
QM-CCC and AOCC-GTO results at and below 50 keV, there are some noticeable 
quantitative differences at higher energies. 
Based on the use of the scaling model \eqref{eq:scale},
transitions to $n>6$ states are insignificant, which
suggests that these discrepancies are likely due to 
numerical issues mentioned earlier
rather than an issue with basis size.

For H($n=2$) collisions, the ionization cross sections
also show a maximum in the intermediate energy region. 
One should note that the cross sections for 
these collisions are larger in magnitude than 
those of H($n=1$) collisions.
It is evident from the comparisons between
the present TC-BGM and the WP-CCC \cite{Abdurakhmanov2018}
results that there are some discrepancies in the vicinity of the maximum.
The results of the scaling model \eqref{eq:scale} in the low-energy region
show that additional states up to $n\approx 7$ are needed
to approximately match the profile of the WP-CCC results.
Note that the WP-CCC \cite{Abdurakhmanov2018} considered 
angular momentum quantum numbers $l\in [0,6]$ and had them assigned 
to $(10-l)$ bound eigenstates on each center.
For example, in the simplest case of $l=0$ there are 10 bound states
on one center in the WP-CCC basis, which is more than the present
basis of 6 bound states for the corresponding $l$ quantum number.
Therefore, the scaling model seems to suggest
that additional bound states (either on the target, projectile, or both)
may be required in the TC-BGM calculation to reach the
same level of agreement with the WP-CCC results
for H($n=2$) collisions as for H($n=1$) collisions.
Moreover, results of the scaling model that involve
much higher shells reduce
the ionization cross sections further. 
Overall, considering that the quantitative
differences between the TC-BGM and WP-CCC results
are no more than approximately 15\%, the results obtained
from the TC-BGM calculations are deemed satisfactory.

\section{Concluding remarks}
\label{sec:end}
The processes of electron excitation, capture,
and ionization in proton collisions with atomic
hydrogen were investigated for the $1s$, $2s$, $2p_{0}$, and $2p_{1}$
initial states.
The present study focuses on collisions at impact energies from 1 to 300 keV.
These processes were quantified by solving
a set of single-particle TDSEs for the expansion coefficients
by using the TC-BGM in the semiclassical approximation.
Based on the close-coupling approach,
the main feature of the TC-BGM is its
construction of dynamically adapted states
which allow for fewer pseudostates in the basis
to reach convergence
compared to the standard two-center AOCC with pseudostates approach.

Overall, the cross sections produced from the present
TC-BGM calculation are in good agreement with previous results.
In particular, the present cross sections  
are mostly similar to the
results based on modern close-coupling implementations such as the
the Sturmian-CC \cite{Winter2009}, 
the recent WP-CCC calculations \cite{Abdurakhmanov2018},
and the AOCC-GTO calculations \cite{Agueny2019}.
The discrepancies between the 
present and previous calculations for ionization can be 
partially attributed to an insufficient number of 
bound states in the present analysis
as shown by the use of a $1/n^{3}$ scaling model on the cross sections.
Notwithstanding these issues, 
the capture and excitation results 
produced in this work can be viewed as a
notable achievement of the TC-BGM
in that a similar level of accuracy as
in other close-coupling approaches is obtained
with a smaller, dynamically-adapted basis.

%
\section{Acknowledgments}
This work was funded by the
Natural Sciences and Engineering Research Council of Canada (NSERC)
and was made possible with the high-performance computing 
resources provided by Compute/Calcul Canada. 
We thank Alain Dubois and Nicolas Sisourat for discussions.

\section{Authors contribution statement}
A.C.K. Leung and T. Kirchner conceived the study, analyzed
the results and drafted the manuscript. 
A.C.K. Leung performed the numerical calculations.
Both the authors have read and approved the final manuscript.
%

\bibliography{ref}

\bibliography{ref}
\bibliographystyle{apsrev4-1}

\end{document}